\documentstyle[preprint,aps]{revtex}
\begin{document}
\preprint{KNUTH-33, ~May ~1996}
\draft

\title{Quantum Fluctuations and Particle Production
of Coherently Oscillating Inflaton}

\author{Jung Kon Kim and Sang Pyo Kim\footnote{
Electronic address: sangkim@knusun1.kunsan.ac.kr}}
\address{Department of Physics\\
Kunsan National University\\
Kunsan 573-701, Korea}

\maketitle
\begin{abstract}
We study a massive inflaton minimally coupled to
the FRW Universe
using the semiclassical gravity
derived from canonical quantum gravity.
It is found that the semiclassical
quantum gravity leads to the power-law expansion
$t^{2/3}$ in an oscillatory phase of the inflaton.
The particle production of the inflaton in the
expanding Universe restricts the duration of stable
coherent oscillations and parametric resonance.
The semiclassical gravity shows
a significant difference that the Hubble constant
does not oscillate in contrast with the oscillatory
behavior in classical gravity.
\end{abstract}
\pacs{04.62.+v, 98.80.Hw}

\section{Introduction}

In the quantum field theory in a curved spacetime
the spacetime is treated as a classical
background geometry, while
matter fields are quantized.
There are two typical approaches to it,
one from conventional field theoretical approach \cite{birrel}
and the other from canonical quantum gravity \cite{kiefer}.
Since the full consideration of quantum fluctuating geometry
and matter field are not still at hand, it would be rather meaningful
to consider the semiclassical gravity obeying the semiclassical
Einstein equation\footnote{We use the units such that
$\hbar = c = k_B = 1$ and $ G = \frac{1}{m_P^2}$.}
\begin{equation}
G_{\mu\nu} = \frac{8 \pi}{m_P^2} \left< \hat{T}_{\mu\nu} \right>,
\end{equation}
where the quantum field
represented by a scalar field $\phi$,
is governed by the time-dependent
Schr\"{o}dinger equation \cite{kiefer}
\begin{equation}
i \frac{\partial}{\partial t}
\Phi (\phi, t) = \hat{H}_m (\phi, t)
\Phi (\phi, t).
\label{sch eq}
\end{equation}
To study a quantum inflaton  from the point of view of
the semiclassical gravity will be of
particular importance and interest in cosmology.

To be more specific, we consider a massive inflaton
minimally coupled to a spatially flat Friedmann-Robertson-Walker
(FRW) Universe
with the metric
\begin{equation}
ds^2 = - N^2 (t) dt^2 + a^2 (t) d \Omega^2_3,
\label{met}
\end{equation}
where $N(t)$ is the lapse function
and $a(t)$ is the scale factor representing
the size of the Universe.
The time-time component classical equation
is the Friedmann equation
\begin{equation}
\left( \frac{\dot{a} (t)}{a (t)} \right)^2 =
\frac{8\pi}{3m_P^2} \frac{T_{00}}{a^3 (t)},
\label{fr eq}
\end{equation}
where
\begin{equation}
T_{00} = a^3 (t) \left(\frac{\dot{\phi}^2 (t)}{2}
+ m^2 \frac{\phi^2 (t)}{2} \right),
\end{equation}
is the energy density of the inflaton.
The classical equation of motion
for the inflaton is
\begin{equation}
\ddot{\phi}(t) + 3 \frac{\dot{a}(t)}{a(t)} \dot{\phi} (t)
+ m^2 \phi (t) = 0.
\label{cl eq}
\end{equation}

This  is the first scheme of our
applying semiclassical gravity
to cosmology.
The main purpose of this paper
is to study the quantum fluctuations and particle production
of a coherently oscillating inflaton,
a homogeneous massive  scalar field minimally coupled to
the spatially flat FRW Universe,
using the recently introduced semiclassical
gravity in which the quantum back
reaction of matter field included \cite{kim1}.
We find exactly the back-reaction originated from the quantum inflaton
which is complex  valued and recognize that it looks  quite  similar as
the classical energy density as a result.
We resort to the WKB-like solution of the quantum inflaton
to study analytically
the semiclassical Einstein gravity equation.
It is shown that the quantum fluctuations
in an oscillatory phase leads to a period of power-law expansion
of matter dominated era until it decays into light bosons
and that the coherently oscillating inflaton suffers
from particle production whose amount is proportional to
the duration of coherent oscillations. The particle production
can restrict the duration of stable coherent oscillations
of the inflaton and affect in a certain way the abundant
(catastrophic) particle
production due to the parametric resonance of bosonic fields
coupled to this coherently oscillating inflaton.

The organization of this paper is as follows.
In Section II, we find
the quantum states of inflaton and
study analytically the semiclassical equation
in the oscillatory phase of inflaton.
In Section III, we discuss the particle creation
due to the quantum fluctuation
of the inflaton.
And in Section IV, we compare the
semiclassical gravity with the classical gravity.

\section{Quantum States of Inflaton}

The massive inflaton in the flat FRW cosmological
model is described as a time-dependent harmonic oscillator
\begin{equation}
H_m = \frac{1}{2 a^3} \pi^2_{\phi} +
\frac{m^2 a^3}{2} \phi^2.
\label{ho}
\end{equation}
The Fock space of the Hamiltonian (\ref{ho}) was
constructed in \cite{kim1}
\begin{equation}
\hat{A}^{\dagger} (t) \hat{A} (t) \left|n, \phi, t\right>
= n \left|n, \phi, t\right>,
\label{fock}
\end{equation}
where
\begin{equation}
\hat{A} (t) = \phi^* (t) \hat{\pi}_{\phi}
- a^3 (t) \dot{\phi}^* (t) \hat{\phi}.
\end{equation}
Furthermore, the expectation value
of the Hamiltonian was given by a simple form
in Ref. \cite{kim2}.
By taking the expectation value with
respect to the number state $\left|n, \phi, t\right> $,
which is a quantum state of Eq. (\ref{sch eq}),
we obtain the semiclassical Einstein equation
\begin{eqnarray}
\left(\frac{\dot{a} (t)}{a (t)} \right)^2 &=&
\frac{8 \pi}{3 m_P^2}
\frac{1}{a^3}
\left<n, \phi,t \right|
\hat{H}_m \left|n, \phi, t \right>
\nonumber\\
&=&
\frac{8 \pi}{3 m_P^2}
\Bigl(n + \frac{1}{2} \Bigr)
\left(\dot{\phi}^* (t) \dot{\phi}(t)
+ m^2 \phi^* (t) \phi (t) \right).
\label{q eq}
\end{eqnarray}
In the above the $\phi$ and $\phi^*$ satisfy
Eq. (\ref{cl eq}) and
the boundary condition
\begin{equation}
a^3 (t) \left ( \dot{{\phi}}^* (t) {\phi}(t)
- {\phi}^* (t) \dot{\phi} (t)
\right) = i.
\label{b c}
\end{equation}
The boundary condition, a Wronskian,
fixes the normalization
constants of two independent solutions.
The boundary condition
is a necessary requirement for the exact
quantum back-reaction at least for
a massive inflaton. It should be remarked that
the semiclassical Einstein equation (\ref{q eq})
has almost the same form as the Friedmann equation
(\ref{fr eq}) except for its complex valued behavior and
that it has the boundary condition entirely determined
by Eq. (\ref{b c})
in strong contrast with the classical theory, in which
the boundary condition for the scalar field is
$\phi (t_0)$ and $\dot{\phi} (t_0)$ at an
initial time $t_0$.

We now solve analytically the self-consistent
semiclassical Einstein equation (\ref{q eq}).
As is well known it is
very difficult to find the classical
solutions of the scalar field in an analytic form,
we rely on the WKB method.
Let the solution have the form
\begin{equation}
\phi (t) = \frac{1}{a^{3/2} (t)}
\varphi (t),
\end{equation}
then
\begin{equation}
\ddot{\varphi} (t) + \left( m^2
- \frac{3}{4} \Bigl(\frac{\dot{a}(t)}{a (t)}\Bigr)^2
- \frac{3}{2} \frac{\ddot{a}(t)}{a(t)}
 \right)
\varphi (t) = 0.
\label{var eq}
\end{equation}
We focus on the oscillatory
phase of inflaton after inflation.
In the parameter region of
\begin{equation}
 m^2> \frac{3 \dot{a}^2}{4a^2}
+ \frac{3 \ddot{a}}{2a},
\label{os con}
\end{equation}
the inflaton has an oscillatory solution
of the form
\begin{equation}
\varphi (t) = \frac{1}{\sqrt{2 \omega (t)}}
\exp \Bigl(- i \int \omega (t) dt \Bigr),
\end{equation}
where
\begin{equation}
\omega^2 (t)
= m^2
- \frac{3}{4} \Bigl( \frac{\dot{a}(t)}{a(t)} \Bigr)^2
- \frac{3}{2} \frac{\ddot{a}(t)}{a(t)}
+ \frac{3}{4} \Bigl(\frac{\dot{\omega} (t)}{\omega (t)} \Bigr)^2
- \frac{1}{2} \frac{\ddot{\omega} (t)}{\omega (t)}.
\label{pha}
\end{equation}
The normalization constant has already been fixed to satisfy
the boundary condition (\ref{b c}).
It should be remarked that the vacuum state in (\ref{fock})
is nothing but the adiabatic vacuum \cite{birrel},
since the WKB-type solutions are used.

Now the semiclassical Einstein equation
can be rewritten as
\begin{equation}
a(t)
= \Biggl[ \frac{4 \pi}{3m_P^2}  \Bigl(n + \frac{1}{2} \Bigr)
\frac{1}{\omega (t) \Bigl(\frac{\dot{a} (t)}{a (t)}
\Bigr)^2 }
\Biggl(m^2 + \omega^2 (t) +  \frac{1}{4} \Bigl(
\frac{\dot{\omega} (t)}{\omega (t)}
+ 3 \frac{\dot{a} (t)}{a(t)} \Bigr)^2 \Biggr)
\Biggr]^{1/3}.
\label{wkb1}
\end{equation}
We solve the Eq. (\ref{wkb1}) perturbatively.
Starting from an approximation ansatz
$\omega_0 (t) = m, ~ a_0 (t) = a_0 t^{2/3}$,
we obtain the next order perturbative
solution
\begin{equation}
a_1 (t)
=
\Biggl[
\frac{6 \pi}{m_P^2} \Bigl(n + \frac{1}{2}
\Bigr) mt^2 \Bigl(1 + \frac{1}{2 m^2 t^2} \Bigr)
\Biggr]^{1/3}
\end{equation}
and
\begin{equation}
\omega_1 (t) = m \Biggl[
 1 + \frac{1}{2m^4 t^4 \Bigl(1 + \frac{1}{2 t^2 m^2} \Bigr)^2}
 \Biggr]^{1/2}.
\end{equation}
The perturbative solutions agree with those in Appendix
obtained from
the fixed background $a(t) = a_0 t^{2/3}$ .
During the later stage ($mt > 1 $) of evolution after
inflation,
the expansion-law is that of the matter dominated era
$a_1 (t) \sim t^{2/3}$ \cite{misner}.
The power-law expansion of the Universe continues
until the inflaton decays into light bosons $\chi$
through interactions of the form $\lambda_1 \phi \chi^2$,
$\lambda_2 \phi^2 \chi^2$, and etc.

\section{Quantum Fluctuation and Particle Production}

The quantum fluctuations of the inflaton
can be found from the dispersion relations
\begin{eqnarray}
\bigl(\Delta \phi \bigr)^2_n &=&
\left<n, \phi, t \right|(\hat{\phi}
- <\hat{\phi}>_n )^2 \left|n,\phi, t \right>
\nonumber\\
&=&  \phi^*_c (t) \phi_c (t)
(2n + 1),
\end{eqnarray}
and
\begin{equation}
\bigl(\Delta \pi_{\phi} \bigr)^2_n
=  a^6 (t) {\dot{\phi}}^*_c (t) \dot{\phi}_c (t)
(2n + 1).
\end{equation}
In the limit $mt >> 1$, the dispersions
become
\begin{eqnarray}
\bigl(\Delta {\phi} \bigr)_n &=&
\sqrt{\frac{1}{a_0^3 m t^2} \Bigl(n + \frac{1}{2} \Bigr)}
\nonumber\\
\bigl(\Delta \pi_{\phi} \bigr)_n &=& \sqrt{
a_0^3 m t^2 \Bigl(1 + \frac{1}{m^2 t^2} \Bigr)
\Bigl(n + \frac{1}{2} \Bigr)}.
\end{eqnarray}
The dispersion of the field decreases
inversely proportionally to
$t$. This means that the quantum field $\hat{\phi}$
becomes strongly peaked along the trajectory,
but it does not describe the coherent state
along the classical trajectory, instead
$<\hat{\phi}> = 0$.
Since the dispersion of momentum
increases proportionally to $t$,
the uncertainty of the quantum state
keeps an asymptotical constant value
\begin{equation}
\bigl(\Delta {\phi} \bigr)_n
\bigl(\Delta \pi_{\phi} \bigr)_n
= \sqrt{1 + \frac{1}{m^2t^2}}
\Bigl(n + \frac{1}{2} \Bigr).
\end{equation}
The uncertainty criterion on the emergence of classical fields
\cite{guth} contrasts with
the common view that the inflaton behaves as a classical
background field after inflation, i.e. in the oscillatory phase.
It has a physical significance
to keep the quantum properties of the oscillating inflaton.

The particle production can be explained
correctly by semiclassical gravity.
The Fock space used in this paper has
a one-parameter dependence on the cosmological time $t$.
The vacuum defined at the initial time $t_0$
contains particles at a later time $t$
\begin{eqnarray}
N_0 (t,t_0) &=&
\left<0, \phi, t_0 \right| \hat{N} (t) \left| 0, \phi, t_0 \right>
\nonumber\\
&=& a^6 |\phi(t) \dot{\phi} (t_0)
- \dot{\phi} (t) \phi (t_0) |^2,
\end{eqnarray}
and the number of particle produced in the $n$th quantum state
is
\begin{equation}
\left<n, \phi, t_0 \right|
\hat{N} (t) \left| n, \phi, t_0 \right>
= n + (2n+1) N_0 (t, t_0).
\end{equation}
In the limit $mt_0, mt > 1$, we use the perturbation
solution to get the number of created particle
\begin{eqnarray}
N_0 ( t, t_0)
&=& \frac{1}{4 \omega(t) \omega(t_0)}
\Bigl(\frac{a(t)}{a(t_0)} \Bigr)^3
\Biggl[\frac{1}{4} \Bigl(3 \frac{\dot{a}(t)}{a(t)}
- 3 \frac{\dot{a}(t_0)}{a(t_0)}
- \frac{\dot{\omega}(t)}{\omega (t)}
+ \frac{\dot{\omega}(t_0)}{\omega (t_0)} \Bigr)^2
+ \bigl(\omega (t) - \omega (t_0) \bigr)^2 \Biggr]
\nonumber\\
&\simeq& \frac{(t - t_0)^2}{4 m^2 t_0^4}.
\end{eqnarray}

The inflaton after inflation can not execute
coherent oscillations for a sufficiently long interval of time
since it suffers from the instability due to the
particle production.
This means
that the inflaton decays into these light bosons
when the inflaton
is coupled to other light bosonic fields,
From the time interval for negligible particle production
\begin{equation}
N_0 (t, t_0) \leq  q
\end{equation}
where $q = {\cal O} (1)$,
we find the duration of stable coherent oscillations
\begin{equation}
\Delta t < \frac{(mt_0)^2 q}{\pi} T,
\label{co os}
\end{equation}
where $T = \frac{2 \pi}{m}$ is the period
of the inflaton oscillation.
The condition (\ref{os con})
for the oscillations immediately after inflation
restricts the interval to the lowest limit
\begin{equation}
\Delta t < \frac{q T}{\sqrt{2} \pi}.
\end{equation}

We discuss effects on the preheating
of the particle production of
the inflaton due to the expansion of the Universe.
The parametric resonance, an abundant particle production
mechanism, requires a sufficient time for the
periodic motion of scalar fields coupled to
the coherently oscillating background inflaton \cite{boyan}.
As explained above the expansion of the Universe restricts
the duration of the stable coherent oscillations
of the inflaton and therefore the amount of particle production
due to the parametric resonance \cite{boyan2}.
The expansion of the Universe also
decreases directly the particle production due
the parametric resonance \cite{kaiser}.
The limit on the duration
of stable coherent oscillations, however, does not
exclude completely the preheating.
We may be able to get a sufficient time for
the stable coherent oscillations
if the oscillatory phase of the inflaton begins
at a later stage $mt_0 >> 1$,
which depends on the inflation models.
Semiclassical gravity treatment of bosonic fields
coupled to the background inflaton will be done
quantitatively in a separate publication.

\section{Comparison with Classical Gravity}

There are several big differences between the classical
and semiclassical gravity approaches.

First of all, the boundary condition in
the semiclassical gravity shows  quite a different
feature from that of the classical gravity.
In the classical gravity one can choose an
arbitrary initial value $\phi_r (t_0)$ and $\dot{\phi}_r (t_0)$.
In the oscillatory phase of the inflaton,
the initial value can be redefined in terms of the amplitude and
phase
\begin{equation}
\phi_r (t) = \frac{{A}_0}{a^{3/2} (t) \sqrt{\omega (t)}}
\sin \bigl(\int \omega (t) dt + \delta \bigr).
\end{equation}
In fact, thanks to the freedom in the choice of the initial values
of the inflaton,  the chaotic inflation models for a broad
class of potentials become possible.
However,  one can not choose arbitrarily the initial values of
the inflaton  any more
in the semiclassical gravity (\ref{q eq}).
Rather, the  amplitude of the inflaton
is fixed by the quantum number $(n + \frac{1}{2})$ and
the time-dependence enters only through the classical solutions
that satisfy the boundary condition (\ref{b c}).
We may choose an arbitrary overall phase factor for the complex
solution since it does not affect
the boundary condition (\ref{b c})
and the semiclassical equation (\ref{q eq}).

Second, the semiclassical equation (\ref{q eq})
does not show any oscillatory behavior even
in the oscillatory phase of the inflaton.
The energy density of quantum state is given by
\begin{equation}
\left<n, \phi, t | \hat{H}_m |n, \phi, t \right>
=  \Bigl(n + \frac{1}{2} \Bigr)
\frac{1}{2 \omega}
\Biggl(m^2 + \omega^2 + \frac{1}{4}
\Bigl(\frac{\dot{\omega}}{\omega} + 3\frac{\dot{a}}{a}
\Bigr)^2
\Biggr)
\end{equation}
On the other hand, the classical energy density
\begin{eqnarray}
H_m = \frac{A_0^2}{2}
\frac{1}{2 \omega} \Biggl[
m^2 + \omega^2 +
\frac{1}{4} \Bigl(\frac{\dot{\omega}}{\omega} + 3\frac{\dot{a}}{a}
\Bigr)^2
+ \Bigl(\omega - \frac{1}{4}
\bigl(\frac{\dot{\omega}}{\omega} + 3\frac{\dot{a}}{a}
\bigr) \Bigr) \cos 2 \bigl(\int \omega dt + \delta \bigr)
\nonumber\\
-
\Bigl(\frac{\dot{\omega}}{\omega} + 3\frac{\dot{a}}{a}
\Bigr)^2
\sin 2 \bigl(\int \omega dt + \delta \bigr),
\Biggr]
\end{eqnarray}
shows manifestly the oscillatory behavior.
As emphasized in Ref. \cite{brand}, the oscillating terms
determine in a significant way the evolution of geometric invariants
through the higher derivatives of the scale factor $a(t)$.
The time average over several oscillations
gives
\begin{equation}
\left< H_m \right>_{t-a} = \frac{A_0^2}{2}
\frac{1}{2 \omega}
\Biggl(m^2 + \omega^2
+ \frac{1}{4}\Bigl(\frac{\dot{\omega}}{\omega} + 3\frac{\dot{a}}{a}
\Bigr)^2
\Biggr)
\end{equation}
assuming that the expansion of the Universe can be neglected during
the time interval.
The expansion of the Universe
due to the time averaged energy density is nearly the same
as that by pressureless dusts.
We may identify the classical amplitude
with $A_0 = \sqrt{2n + 1}$.

\section{Conclusion}

In this paper we studied analytically
the semiclassical gravity of
a massive inflaton minimally coupled to
the spatially flat Friedmann-Robertson-Walker
Universe.
It was found that the back-reaction of the quantum inflaton
obeying the time-dependent
Schr\"{o}dinger equation leads to the power-law
expansion $t^{2/3}$ of the Universe in the oscillatory phase.
The power-law expansion is the same as that of
the matter (dust particles) dominated era, but
not $t^{1/2}$ of relativistic particles.
The criterion based on the uncertainty
shows that the oscillating inflaton after inflation
is a quantum field rather than a classical one.

The expansion of the Universe driven by
the inflaton causes in turn the particle production of the
inflaton itself. The particle production can restrict
in a certain way the
time interval of stable coherent oscillations.
Thus due to the particle production of the inflaton
originating from the expansion of the
Universe, the preheating mechanism is subject to
the limited duration of stable coherent
oscillations.
This, however, does not mean the complete exclusion of
the preheating as a mechanism for the abundant
particle and entropy production needed for the
formation of the present Universe.
The number of particles produced
from the light bosons coupled to the inflaton
executing coherent oscillations for a limited
period shows a power-law increase rather than
an exponential increase in the case of the non-decaying
inflaton oscillating stably for an infinite
period in a static universe.

We also discussed the difference of the boundary
conditions between semiclassical and classical
gravity. One prominent difference is that
the semiclassical gravity does not show
any oscillatory behavior of the Hubble constant
in strong contrast with the oscillatory behavior
of classical gravity. Furthermore,
the solution of the inflaton in semiclassical gravity
is fixed by the quantum number up to an overall phase
factor.

\acknowledgements
We thank  Dr. J. Y. Ji for useful discussions.
This work was supported by
the Korea Science and Engineering Foundation
under Grant No. 951-0207-056-2.

\appendix
\section{Quantum Inflaton in a Fixed
Background Geometry}

We find the exact quantum states of the inflaton
assuming that the expansion of the Universe
is fixed by the power-law $a (t) = a_0 t^{2/3}$.
The classical field equation of the inflaton
\begin{equation}
\ddot{\phi}_b (t) + \frac{2}{t} \dot{\phi}_b (t)
+ m^2 \phi_b (t) = 0,
\end{equation}
has a complex solution
\begin{equation}
\phi_b (t) = \sqrt{\frac{\pi}{4 a_0^3 t}}
H_{-1/2}^{(2)} (mt).
\end{equation}
The complex solution satisfies the boundary
condition (\ref{b c}).
Following Ref. \cite{kim2}, we construct the Fock space
of the inflaton in terms of the creation  and
annihilation operator
\begin{eqnarray}
\hat{A}^{\dagger} (t) = \phi_b (t) \hat{\pi}_{\phi}
- a_0^3 t^2  \dot{\phi}_b (t) \hat{\phi},
\nonumber\\
\hat{A} (t) = \phi^*_b (t) \hat{\pi}_{\phi}
- a_0^3 t^2 \dot{\phi}^*_b (t) \hat{\phi}.
\end{eqnarray}
For $mt >> 1$ we obtain the asymptotic form
\begin{equation}
\phi_b (t) = \sqrt{\frac{1}{2 a_0^3 m}}
\frac{1}{t} e^{- imt},
\end{equation}
which has the same form as the WKB solution
obtained without fixing the background geometry.

\end{document}